\documentclass[aps,pra,groupedaddress,superscriptaddress,nofootinbib]{revtex4}

\usepackage{graphicx,amsmath,amssymb,amsbsy,subfigure,hyperref,bbm,times,txfonts,float}
\usepackage{subfigure,hyperref,bbm,times}
\usepackage[T1]{fontenc}
\usepackage{braket}
\usepackage{epsfig} 
\usepackage{color}
\usepackage{xcolor}%
\usepackage{graphicx}
\usepackage{dcolumn}
\usepackage{bm}

\providecommand{\openone}{\leavevmode\hbox{\small1\kern-3.8pt\normalsize1}}

\usepackage{soul}

\hypersetup{
	colorlinks=true,
	linkcolor=blue,
}
\graphicspath{{figure/}}

\begin{document}

	\title{Synergistic Effects of Detuning and Auxiliary Qubits on Quantum Synchronization}
	\author{Amir Hossein Houshmand Almani}
	\affiliation{Department of Physics, University of Guilan, P. O. Box 41335-1914, Rasht, Iran}
	\author{Ali Mortezapour}
	\email{mortezapour@guilan.ac.ir}
	\affiliation{Department of Physics, University of Guilan, P. O. Box 41335-1914, Rasht, Iran}
	\author{Alireza Nourmandipour}
	\affiliation{Department of Physics, Sirjan University of Technology, 7813733385 Sirjan, Iran}
	
	\begin{abstract}

		We investigate how detuning and auxiliary qubits collaboratively enhance quantum synchronization in a dissipative multi-qubit system that is coupled to a structured reservoir. Our findings indicate that while detuning is ineffective in Markovian environments, it emerges as a powerful control parameter in the non-Markovian regime, where environmental memory facilitates long-lived phase coherence. It is shown that adding more auxiliary qubits amplifies this effect by strengthening the collective coupling and enhancing memory, resulting in robust phase locking within the system. Analysis using the Husimi Q-function, synchronization measures, and Arnold tongue structures reveals a detuning-induced enhancement of phase locking, which significantly improves stability compared to the resonance case. These results establish a cooperative control strategy where detuning actively engineers phases, while auxiliary qubits provide the necessary memory for sustained synchronization.
	\end{abstract}
	
	\date{\today }
	
	\maketitle

	\section{INTRODUCTION}
Synchronization represents one of the most fundamental collective phenomena observed throughout nature, governing processes ranging from the coordinated firing of neurons to the synchronized flashing of firefly colonies. This area of study has deep historical roots, tracing back to 1673 when Huygens first documented the coordinated motion of two weakly coupled pendulums \cite{1}. In classical systems, synchronization occurs when coupled oscillators—whether mechanical, chemical, or biological—adjust their rhythms through mutual interaction or external driving, resulting in frequency and phase locking. The study of synchronization has deep roots in nonlinear dynamics, with pioneering work on models such as the van der Pol oscillator providing crucial insights into the mechanisms underlying this behavior \cite{2,3,4}.

In recent decades, the exploration of synchronization has expanded into the quantum domain, where it takes on new dimensions due to the principles of quantum mechanics. Quantum synchronization differs fundamentally from its classical counterpart, as it arises from the interplay of coherence, entanglement, and the probabilistic nature of quantum states \cite{5,6,7,8,9}. While synchronization in classical systems is often associated with the emergence of limit cycles, where periodic oscillations lead to phase locking between coupled oscillators, synchronization in quantum systems can arise through different mechanisms, even without such oscillatory behavior \cite{7}.
 
The quantum analog of synchronization is predominantly investigated within open quantum systems, as isolated quantum systems typically exhibit complex eigenfrequency spectra with arbitrary phase relationships, leading to irregular, poorly defined observable oscillations \cite{8,11}. Synchronization is generally divided into two categories: forced synchronization and spontaneous (or mutual) synchronization. Spontaneous synchronization occurs during the transient dynamics of a system, arising naturally from interactions among its components or with the surrounding environment \cite{13,14,15,16,17,18,19,20,21}.  In contrast, forced synchronization, also known as entrainment, results from the influence of external driving forces \cite{22,23,24,25,26,27,28}. In the context of open quantum systems, synchronization most often occurs during relaxation processes, a form of spontaneous synchronization in which the system dissipates energy as it interacts with its surroundings. Recent studies have concentrated on synchronization in finite-dimensional quantum systems, including qubits \cite{29,30,31,32} and qutrits \cite{18,24,25}. Notable experimental advancements, such as the synchronization of a single trapped-ion qubit, have validated these theoretical predictions and opened new avenues for research \cite{33}.
The synchronization of quantum systems offers significant potential for applications in quantum technologies, including quantum communication \cite{34}, metrology \cite{35}, and the development of scalable quantum networks \cite{33,37}. For example, synchronized quantum systems could improve sensor precision and facilitate robust quantum information transfer within networks.

Recent research has demonstrated that introducing auxiliary qubits into dissipative quantum systems can serve as an effective strategy for preserving quantum resources. In particular, studies have explored $N$-qubit systems coupled to zero-temperature reservoirs, where the strategic incorporation of additional qubits has proven beneficial for protecting quantum resources. This approach has shown remarkable success across a range of quantum phenomena. Researchers have established that auxiliary qubits can effectively safeguard entanglement in both two-party and multi-party quantum systems \cite{40,41,42,43,44}. Similarly, quantum coherence \cite{45,46,47,48} and the entropic uncertainty principle lower bound \cite{49} have been successfully protected by this method within generalized amplitude-damping frameworks. The technique has also demonstrated its utility in preserving quantum discord \cite{50} and entropy squeezing \cite{51}. A particularly noteworthy application emerged in quantum battery research, where auxiliary qubits were found to significantly suppress the self-discharging mechanisms that typically degrade quantum battery performance \cite{52}. This finding suggests that the auxiliary qubit approach may have broader implications for quantum energy storage systems.

Motivated by these considerations, this paper proposes a mechanism for controllably synchronizing the phase of a single-qubit system. This approach relies on manipulating the system's reservoir by adding non-interacting qubits and detuning. We consider a system of $N$ qubits coupled to a common dissipative reservoir. It is noteworthy that a closely related study examined quantum synchronization under strict resonance and found that auxiliary qubits alone can enhance the Husimi Q-function and synchronization measure, yielding long-time phase preference that depends solely on the number of added qubits \cite{55}. In contrast, the present work incorporates non-zero detuning ($\Delta \neq 0$) as an active control parameter. It demonstrates that detuning fundamentally modifies the synchronization dynamics, enabling robust, long-lived phase locking even in regimes where resonance-based schemes lose coherence. This distinction highlights that detuning and auxiliary qubits play complementary roles, transforming phase protection from a purely resource-driven effect into an actively tunable synchronization mechanism. 

The paper is organized as follows: we give an overview of the model Hamiltonian and the qubit dynamics in Sec. \ref{model}. In Sec. \ref{sec.RD}, we investigate in detail the Husimi $Q$-function and a measure to identify effective factors in the system to achieve phase locking. We then turn to the Arnold torque and Bloch sphere parameters of the system to verify the existence of quantum phase synchronization. Finally, we present an outline of the main findings and prospects in Sec. \ref{sec.iv}.

	\section{The Model}
\label{model}
	
	We consider a quantum system consisting of $N$ identical, non-interacting qubits that are collectively coupled to a common zero-temperature bosonic reservoir. This configuration represents a natural extension of the well-known damped Jaynes-Cummings model, where multiple qubits share the same environmental coupling rather than a single qubit interacting with the reservoir.
	
	Under the rotating-wave approximation, the total Hamiltonian of our system can be expressed as:
	\begin{equation}
		H = H_0 + H_I \quad \text{(with } \hbar = 1\text{)},
	\end{equation}
	where the free Hamiltonian is:
	\begin{equation}
		H_0 = \omega_0 \sum_{i=1}^{N} \sigma_i^+ \sigma_i^- + \sum_k \omega_k b_k^\dagger b_k,
	\end{equation}
	and the interaction Hamiltonian reads:
	\begin{equation}
		H_I = \sum_{i=1}^{N} \sum_k \left( g_k \sigma_i^+ b_k + g_k^* \sigma_i^- b_k^\dagger \right).
	\end{equation}
	
	Here, $\sigma_i^+ = \ket{1}_i\bra{0}$ and $\sigma_i^- = \ket{0}_i\bra{1}$ represent the raising and lowering operators for the $i$-th qubit, respectively. All qubits share the same transition frequency $\omega_0$, while $b_k^\dagger$ and $b_k$ denote the creation and annihilation operators for the $k$-th reservoir mode with frequency $\omega_k$. The coupling strength $g_k$ between each qubit and reservoir mode $k$ is assumed to be identical for all qubits.
	
	To obtain an exact analytical solution, we restrict our analysis to the single-excitation subspace of the total system. This constraint is physically motivated by the conservation of the total excitation number:
	\begin{equation}
		\left[ \sum_{i=1}^{N} \sigma_i^+ \sigma_i^- + \sum_k b_k^\dagger b_k, H \right] = 0.
	\end{equation}
	
	The most general initial state within this subspace can be written as:
	\begin{equation}
		\ket{\Psi(0)} = C_0(0)\ket{0}_S\ket{0}_R + \sum_{i=1}^{N} C_i(0)\ket{i}_S\ket{0}_R,
	\end{equation}
	where $\ket{0}_S$ represents the state with all qubits in their ground states, $\ket{i}_S$ denotes the state where only the $i$-th qubit is excited, and $\ket{0}_R$ is the vacuum state of the reservoir.
	
	The time-evolved state maintains the single-excitation constraint and takes the form:
	\begin{equation}
		\ket{\Psi(t)} = C_0(0)\ket{0}_S\ket{0}_R + \sum_{i=1}^{N} C_i(t)\ket{i}_S\ket{0}_R + \sum_k G_k(t)\ket{0}_S\ket{1_k}_R,
	\end{equation}
	where $\ket{1_k}_R$ indicates a single excitation in the $k$-th reservoir mode.
	
	Applying the Schrödinger equation in the interaction picture and using standard procedures, we derive the coupled differential equations:
	\begin{align}
		\dot{C}_i(t) &= -i \sum_k g_k e^{-i(\omega_k - \omega_0)t} G_k(t), \\
		\dot{G}_k(t) &= -i g_k^* e^{i(\omega_k - \omega_0)t} \sum_{i=1}^{N} C_i(t),
	\end{align}
	
	Utilizing the initial condition $G_k(0) = 0$ and eliminating the reservoir degrees of freedom, we obtain the integro-differential equations:
	\begin{equation}
		\dot{C}_i(t) = -\int_0^t f(t-t') \sum_{j=1}^{N} C_j(t') dt' \quad (i = 1, 2, \ldots, N),
	\end{equation}
	
	The memory kernel $f(t-t')$ encodes the environmental correlations and is related to the reservoir spectral density $J(\omega) = \sum_k |g_k|^2 \delta(\omega - \omega_k)$ through:
	\begin{equation}
		f(t-t') = \int_{-\infty}^{\infty} d\omega \, J(\omega) e^{i(\omega_0 - \omega)(t-t')}.
	\end{equation}
	
	We adopt a Lorentzian spectral density to model the structured reservoir:
	\begin{equation}
		J(\omega) = \frac{1}{2} \gamma_0 \frac{\lambda^2}{(\omega_0 - \Delta - \omega)^2 + \lambda^2},
	\end{equation}
	where $\Delta = \omega_0 - \omega_c$ is the detuning between the qubit frequency and the reservoir central frequency $\omega_c$. The parameter $\lambda$ characterizes the spectral width and is inversely related to the reservoir correlation time ($\tau_B = \lambda^{-1}$), while $\gamma_0$ determines the relaxation time scale ($\tau_R = \gamma_0^{-1}$).
	
	Through Laplace transform techniques, the exact solution for the probability amplitudes is:
	\begin{align}
		C_i(t) &= e^{-dt/2} \left[ \cosh\left(\frac{Dt}{2}\right) + \frac{d}{D} \sinh\left(\frac{Dt}{2}\right) \right] C_i(0) \nonumber \\
		&\quad + \frac{(N-1)C_i(0) - \sum_{j \neq i} C_j(0)}{N} \left\{ 1 - e^{-dt/2} \left[ \cosh\left(\frac{Dt}{2}\right) + \frac{d}{D} \sinh\left(\frac{Dt}{2}\right) \right] \right\},
	\end{align}
	where we have defined $d = \lambda - i\Delta$ and $D = \sqrt{d^2 - 2N\gamma_0\lambda}$.
	
	For the physically relevant case where initially only one qubit (say, the first) is in the excited state ($C_j(0) = 0$ for $j \neq i$), the solution simplifies to:
	\begin{equation}
		C_i(t) = C_i(0) h(t),
	\end{equation}
	where the decay function is:
	\begin{equation}
		h(t) = \frac{N-1}{N} + \frac{1}{N} e^{-dt/2} \left[ \cosh\left(\frac{Dt}{2}\right) + \frac{d}{D} \sinh\left(\frac{Dt}{2}\right) \right]
		\label{Eq14}.
	\end{equation}
	
	This result reduces to the standard damped Jaynes-Cummings solution when $N = 1$. Crucially, the Markovian and non-Markovian regimes are now determined by the conditions $\lambda > 2N\gamma_0$ and $\lambda < 2N\gamma_0$, respectively, showing that the presence of auxiliary qubits fundamentally alters the system's dynamical characteristics.
	
	The reduced density matrix of the $i$-th qubit in the computational basis $\{\ket{1}_i, \ket{0}_i\}$ is given by:
	\begin{equation}
		\rho_i(t) = \begin{pmatrix} 
			\rho_{11}^i(0)|h(t)|^2 & \rho_{10}^i(0)h(t) \\
			\rho_{01}^i(0)h^*(t) & 1 - \rho_{11}^i(0)|h(t)|^2 
		\end{pmatrix},
		\label{Eq15}
	\end{equation}
	where $\rho_{11}^i(0) = |C_i(0)|^2$ and $\rho_{10}^i(0) = \rho_{01}^{i*}(0) = C_i(0)C_0^*(0)$ are determined by the initial conditions.
	
	This formulation provides the foundation for analyzing quantum phase synchronization through various measures, as will be discussed in the following sections.

\section{Results and discussions}
\label{sec.RD}

\subsection{Husimi function and synchronization}

The Husimi $Q$-function represents a quasi-probability distribution in phase space that provides a smoothed, non-negative representation of a quantum state's density matrix. This powerful visualization and analysis tool bridges classical phase space concepts of position and momentum with quantum mechanical principles, making it invaluable for studying quantum states.
In quantum synchronization research, the $Q$-function serves as a crucial tool for demonstrating phase coherence and phase locking phenomena—fundamental indicators of synchronization behavior. Under non-synchronized conditions, the $Q$-function exhibits a nearly uniform phase distribution, reflecting the absence of any preferred phase orientation. Conversely, when synchronization occurs, a distinct peak emerges at a specific phase location, indicating phase locking where the system becomes dynamically attracted to a particular phase state.
The temporal persistence of this peak demonstrates that the system has achieved a stable phase configuration. In our theoretical framework, the qubit's frequency modulation governs its interaction with the dissipative cavity environment, resulting in dynamic modifications of the Husimi $Q$-function. The emergence and stabilization of phase peaks provide unambiguous signatures of synchronization phenomena.
As a quasi-probability distribution, the Husimi $Q$-function effectively bridges classical and quantum mechanical descriptions, providing valuable insights into synchronization emergence in quantum systems while reducing the mathematical complexity inherent in alternative phase-space representations like the Wigner function. This characteristic makes the $Q$-function particularly suitable for investigating classical-to-quantum transitions and examining how decoherence and noise affect synchronization processes.
Through $Q$-function analysis, synchronization phenomena can be identified, monitored, and characterized, thereby enhancing our understanding of coupled quantum system dynamics.
For qubit systems, the Husimi $Q$-function is formally expressed as:
\begin{equation}
	Q(\theta, \phi, t) = \frac{1}{2\pi} \langle\theta, \phi| \rho(t) |\theta, \phi\rangle,
\end{equation}
where the states $|\theta, \phi\rangle = \cos\frac{\theta}{2} |e\rangle + \sin\frac{\theta}{2} e^{i\phi} |g\rangle$ represent the eigenstates of the operator $\hat{\sigma} \cdot \mathbf{n}$ with $\mathbf{n} = (\sin\theta \cos\phi, \sin\theta \sin\phi, \cos\theta)$.

The explicit form of the Husimi Q-function can be straightforwardly derived:
\begin{align}
	Q(\theta, \phi, t) &= \frac{1}{2\pi} \left[ \rho_{11}(t) \cos^2\frac{\theta}{2} + \rho_{00}(t) \sin^2\frac{\theta}{2} \right. \nonumber \\
	&\quad \left. + \frac{1}{2}\sin\theta \left( \rho_{10}(t) e^{i\phi} + \rho_{01}(t) e^{-i\phi} \right) \right].
\end{align}

This formulation enables precise calculation of phase distributions and quantitative analysis of synchronization behavior.

\begin{figure}[!htbp]
	\centering
	\includegraphics[width=0.9\textwidth]{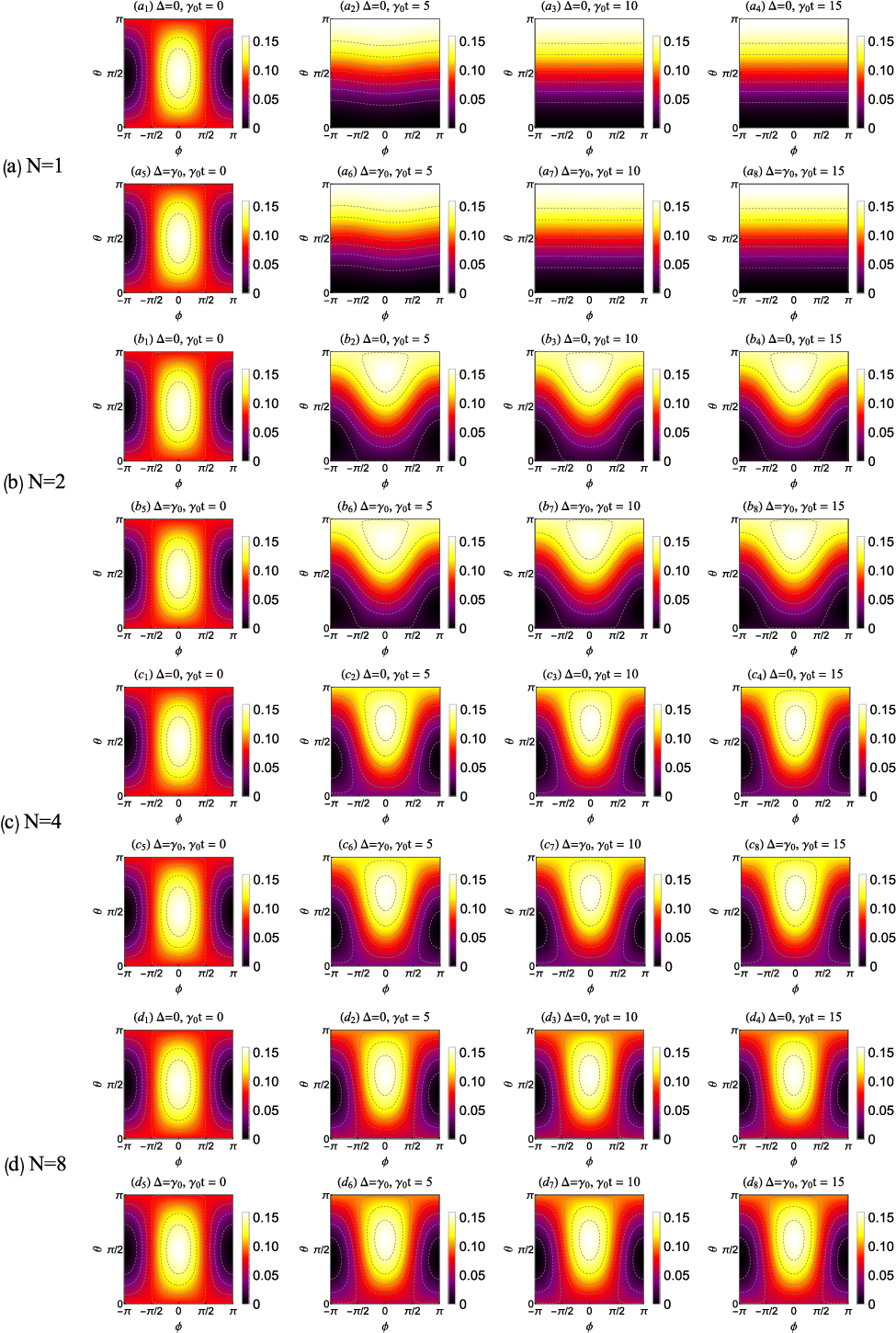}	
	\caption{Husimi $ Q $-function $ Q(\theta, \phi, t) $ for various numbers of ancillary qubits ($N$) in the Markovian regime ($\lambda = 5\gamma_0$). The top row shows resonance condition ($\Delta = 0$), while the bottom row correspond to $\Delta = \gamma_0$.} 
	\label{Fig1}
\end{figure}

\begin{figure}[!htbp]
	\centering
	\includegraphics[width=0.9\textwidth]{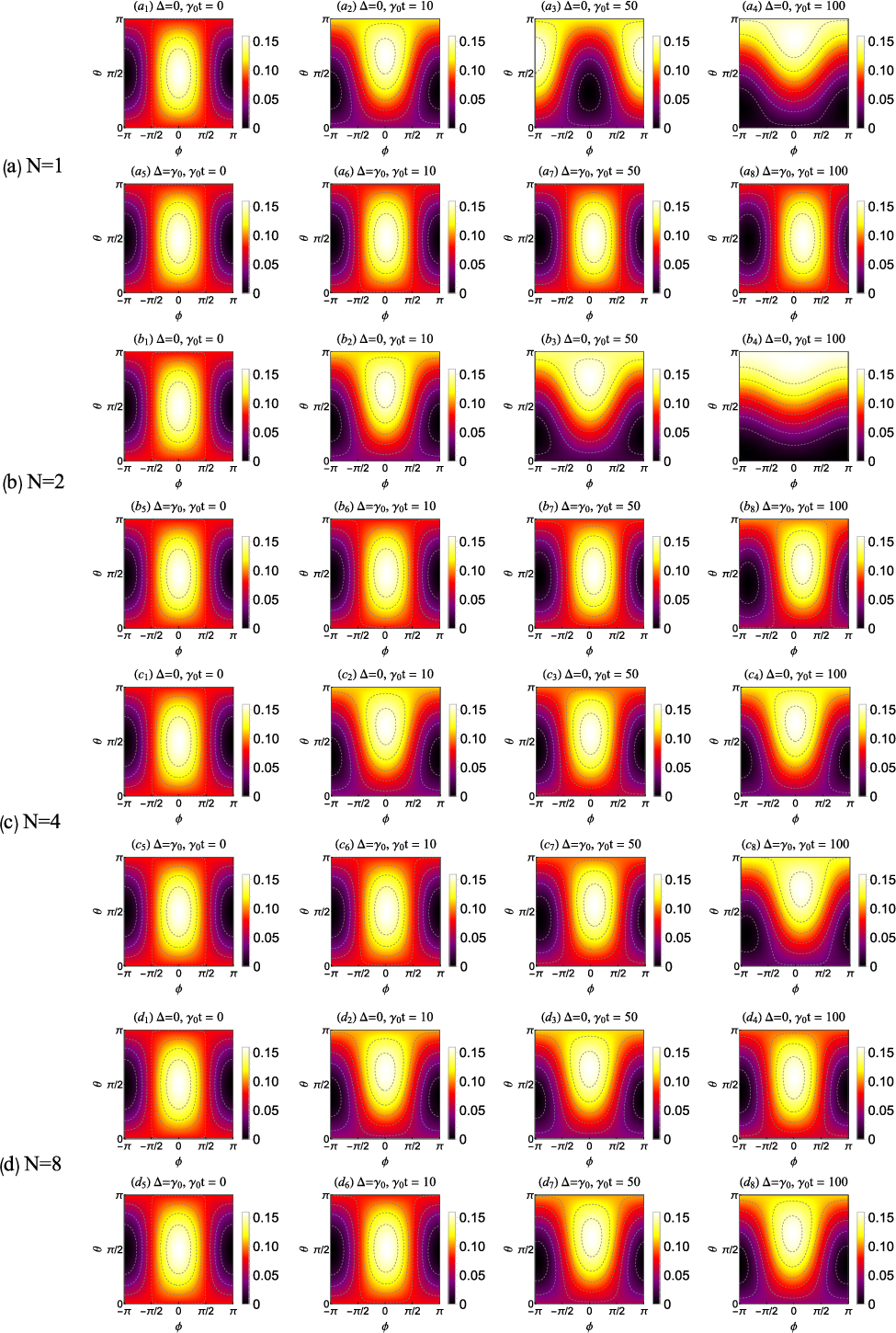}	
	\caption{Husimi $ Q $-function $ Q(\theta, \phi, t) $ for various numbers of ancillary qubits ($N$) in the non-Markovian regime ($\lambda = 0.01\gamma_0$). The top row shows resonance condition ($\Delta = 0$), while the bottom row correspond to $\Delta = \gamma_0$.} 
	\label{Fig2}
\end{figure}

Figure \ref{Fig1} shows the evolution of the Husimi $Q$-function as a function of $\theta$ and $\phi$ within the Markovian regime ($\lambda = 5\gamma$). We compare zero detuning (top row) with non-zero detuning (bottom row) as the number of auxiliary qubits is varied. In the absence of detuning ($\Delta = 0$; top row), the $Q$-function rapidly transitions from an initially peaked distribution at $\theta = \pi/2$, $\phi = 0$ towards uniformity, indicating irreversible dephasing. The addition of auxiliary qubits (increasing $N$) slightly mitigates this decay by populating the decoherence-free subspace, evident in the $N^{-1}$ weighting of the decaying component in $h(t)$. However, it ultimately cannot avert phase randomization. Notably, introducing finite detuning ($\Delta = \gamma_0$ bottom row) results in minimal qualitative change: the system still converges to a uniform distribution. This insensitivity arises because, in the Markovian limit, the memory kernel $f(t-t')$ behaves like a delta function, leading to a local-in-time master equation in which $\Delta$ only induces a trivial rotation about the $z$-axis. Lacking the memory to retain phase information, this rotation cannot maintain phase locking in the face of persistent dissipative loss. The environment acts as an infinite sink, rendering detuning an ineffective control parameter. It is also noteworthy that our findings for $\Delta = 0$ are consistent with those obtained by et al \cite{55}.

In Figure \ref{Fig2}, we examine the evolution of the Husimi $Q$-function in the non-Markovian regime ($\lambda = 0.01\gamma_0$). It is observed that in the resonance condition ($\Delta = 0$: top row), the oscillatory behavior in the $Q$ function is restored due to the feedback flow of information, due to the memory effect. This is due to the non-exponential nature of $h(t)$. However, these oscillatory revivals are temporary and tend to fade over time unless the number of particles, $N$, is large ($N = 8$). In such cases, the collective coupling $N\gamma_0 \approx 0.08\gamma_0$ is sufficient to surpass $\lambda$, thereby enhancing coherence through the reinforcement of non-Markovian dynamics. The breakthrough happens when detuning is introduced ($\Delta = \gamma_0$; bottom row). In fact, detuning creates a coherent phase rotation rate that competes with the lossy and recursive dynamics dictated by the memory core. In mathematical terms, this competition causes the expression $d = \lambda - i\Delta$ to become complex. As a result, the evolutionarily damped function $h(t)$ becomes a function with damped oscillations whose frequency is determined by $\Delta$. When these oscillations match the memory feedback timescale $\sim \lambda^{-1}$, they interfere constructively to stabilize the phase and reduce incoherence. Such phase locking due to detuning, even with minimal resources ($N = 2$), produces a robust and stable peak in the $Q$ function, making the $N = 2$ dynamics almost identical to the $N = 8$ dynamics. Note that the key here is the role of memory in temporarily storing phase information that detuning can later retrieve and lock, rather than allowing it to leak irreversibly.

\subsection{{Synchronization Measure $S(\phi,t)$}}

To quantitatively assess the degree of synchronization in our quantum system, we introduce a synchronization measure. This measure provides a rigorous framework for evaluating how closely the phases of the qubit align with those of its environment over time. By examining this synchronization measure, we can systematically analyze the impact of varying qubit velocity on synchronization dynamics. In this section, we will define and calculate the synchronization measure, explore its theoretical foundations, and present its application within the context of our model. By using this measure, we aim to gain precise insights into the mechanisms that drive phase locking and coherence preservation, which are essential for the effective implementation of quantum synchronization in practical quantum technologies. To achieve this, we express the synchronization function $S(\phi, t)$ by integrating the function $Q(\theta, \phi, t)$ with respect to $\theta$ as follows:
\begin{equation}
S(\phi, t) = \int_{0}^{\pi} d\theta \sin\theta Q(\theta, \phi, t) - \frac{1}{2\pi}.
\label{Eq18}
\end{equation}
By evaluating the integral over the angular variable $\theta$ in Eq. (\ref{Eq18}) and applying the trace invariance condition, $\rho_{00}(t) + \rho_{11}(t) = 1$, we obtain
\begin{equation}
	S(\phi, t) = \frac{1}{8} |h(t)| \cos\left[ \phi + \arctan\left( \frac{\operatorname{Im}(h(t))}{\operatorname{Re}(h(t))} \right) \right].
\end{equation}
This measure isolates the phase-coherent component from population dynamics, with $S \to 0$ indicating complete desynchronization and $|S| > 0$ signaling phase-locking (positive for in-phase, negative for anti-phase).
\\

\begin{figure}[!htbp]
	\centering
	\includegraphics[width=1\textwidth]{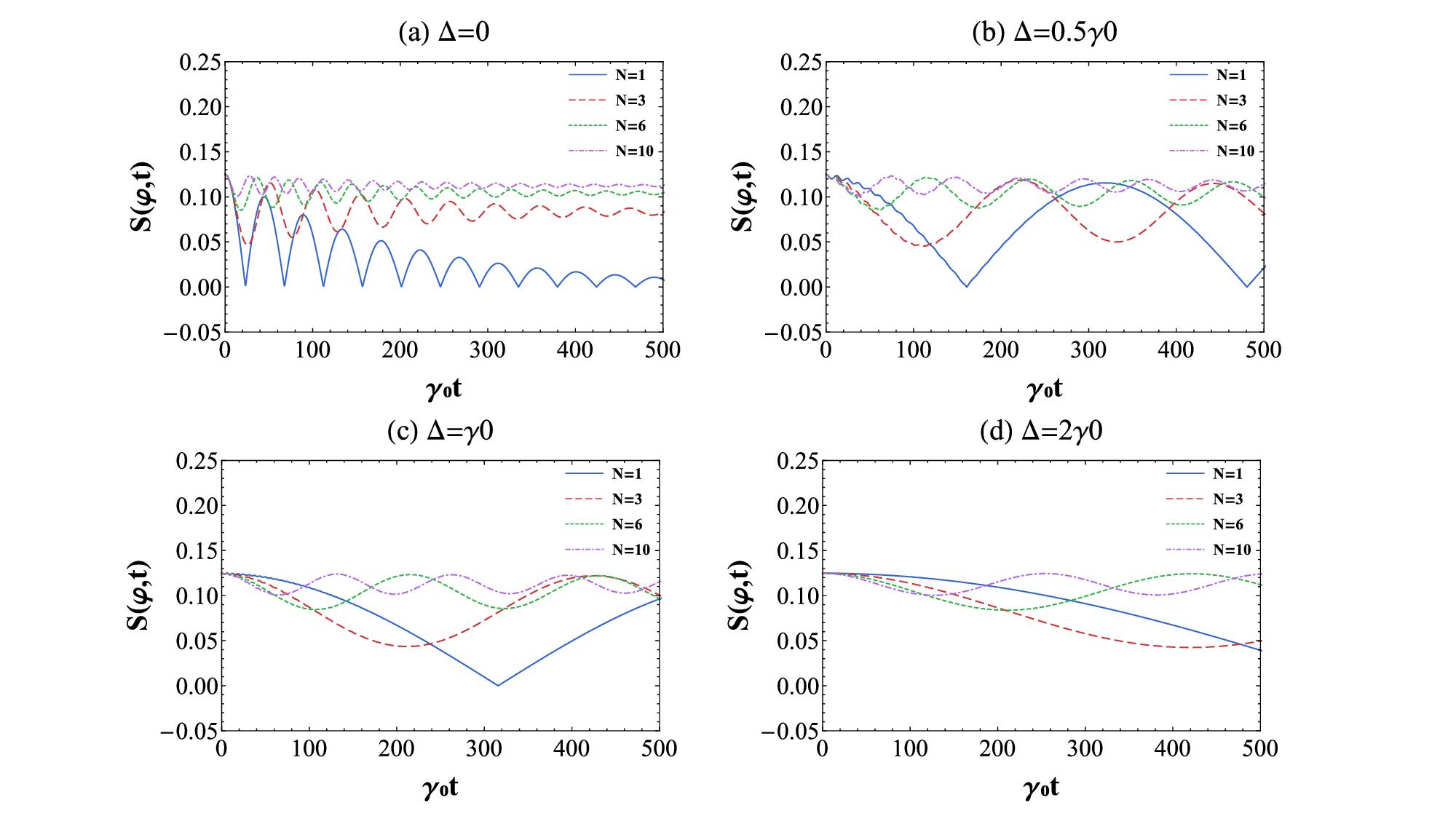}	
	\caption{Synchronization measure $ S(\phi=0,t) $ in the non-Markovian regime ( $\lambda = 0.01\gamma_0$ ) and several values of $N$ and $\Delta = 0$. Panels corresponds to (a) $\Delta = 0$ , (b) $\Delta = 0.5\gamma_0$, (c) $\Delta = \gamma_0$, and (d) $\Delta = 2\gamma_0$.} 
	\label{Fig3}
\end{figure}

Figure \ref{Fig3} displays the synchronization measure $S(\phi = 0, t)$ for various values of $\Delta$ and numbers of auxiliary qubits in the non-Markovian regime ($\lambda = 0.01\gamma_0$). It can be observed that for $\Delta = 0$ (see Fig. \ref{Fig3}(a)), $S(\phi = 0, t)$ shows pronounced damped oscillations when the number of auxiliary qubits $N$ is small ($N = 1, 3$). These oscillations diminish gradually as $N$ increases. Note that these oscillations arise from the competition between the induced memory recovery and the intrinsic decoherence rate that continuously leaks phase information. This competition leads to unstable phase slips and prevents steady-state locking unless $N$ is sufficiently large to induce a quantum Zeno-like suppression of the oscillations. In contrast, Fig. \ref{Fig3}(b) - Fig. \ref{Fig3}(d) (panels (b–d)) disclose that even modest detuning ($\Delta \ge 0.5\gamma_0$) eliminates these oscillations, driving $S$ to a constant non-zero value for all $N$. The transient period lengthens with increasing $\Delta$, as evidenced by the progressively slower convergence to steady state across panels (b–d). This period lengthening reflects the fact that detuning modifies the effective eigenfrequencies of the non-Markovian dynamics, stretching the memory-feedback timescale and slowing the phase-locking convergence. However, for any fixed $\Delta \neq 0$, the synchronized value of $S$ is nearly independent of $N$, indicating that detuning provides robust phase-locking without requiring large numbers of auxiliary qubits.

\subsection{{Arnold Tongue Structures in Parameter Space}}

In classical synchronization theory, the Arnold tongue typically defines regions of synchronization in the plane formed by external drive frequency versus coupling strength. In these regions, the phase of an oscillator becomes locked to an external drive. Recently, this concept has been applied to open quantum systems, particularly qubits, by mapping its analog in the same plane as a function of detuning and coupling strength \cite{32,27}. In this context, detuning represents an intrinsic rotation frequency, while coupling facilitates dissipation. To find this quantum analog, one can compute $S(\phi, t)$ over the entire range of $\phi$ and identify the maximum value. The expression for the maximum shifted phase distribution, denoted as $S_m(t)$, is defined as follows:

\begin{equation}
	S_m(t) = \frac{1}{8}|h(t)|.
\end{equation}

\begin{figure}[!htbp]
	\centering
	\includegraphics[width=1\textwidth]{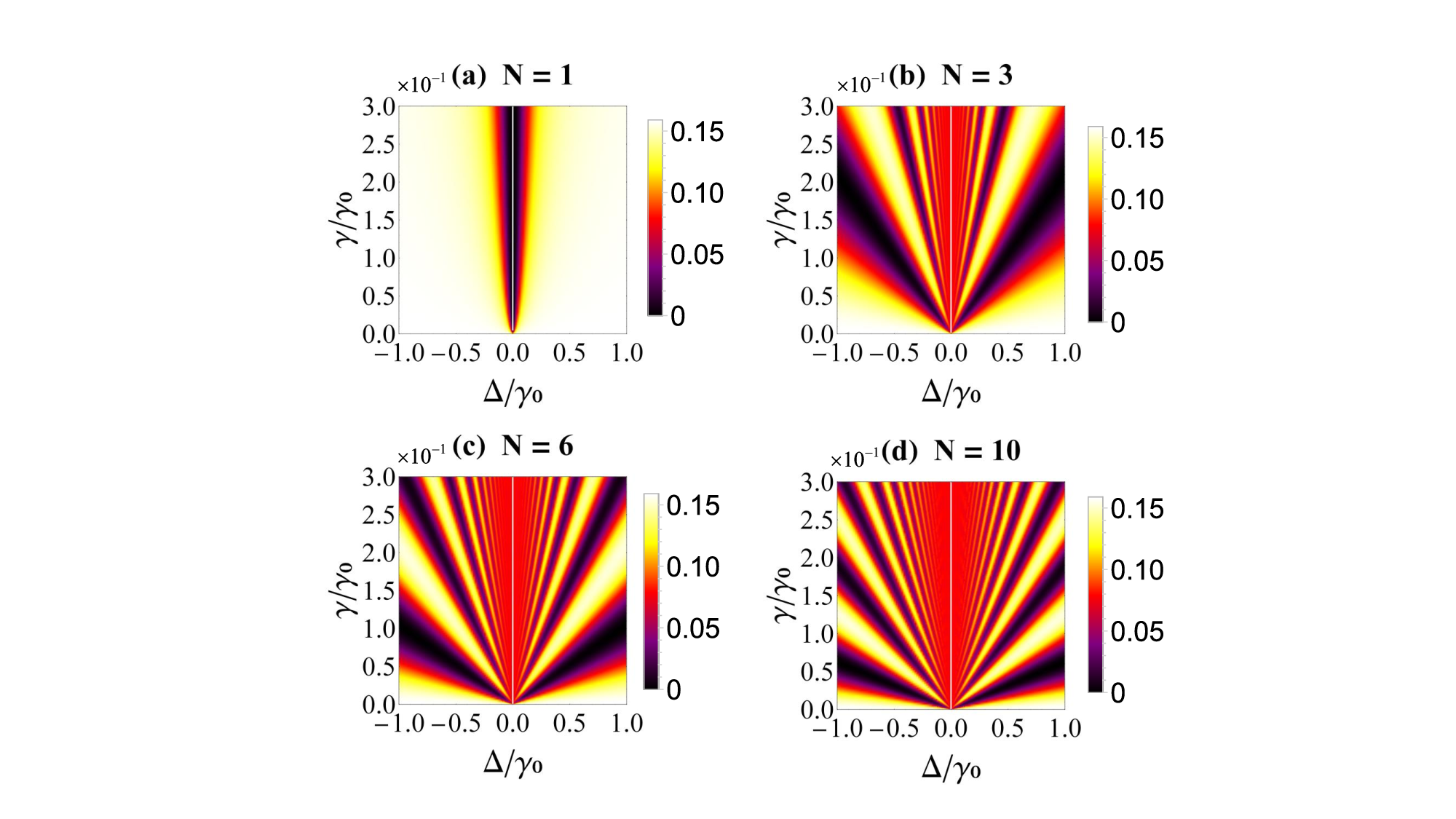}		
	\caption{Maximum synchronization measure $S_m(t)$ as a function of the detuning $\Delta$ and coupling strength $\gamma$ (in units of $\gamma_0$) at time $\gamma_0t=1000$ for $\lambda=0.01\gamma_0$. Panels show (a) $N=1$, (b) $N=3$, (c) $N=6$, and (d) $N=10$.} 
	\label{Fig4}
\end{figure}
\begin{figure}[!htbp]
	\centering
	\includegraphics[width=1\textwidth]{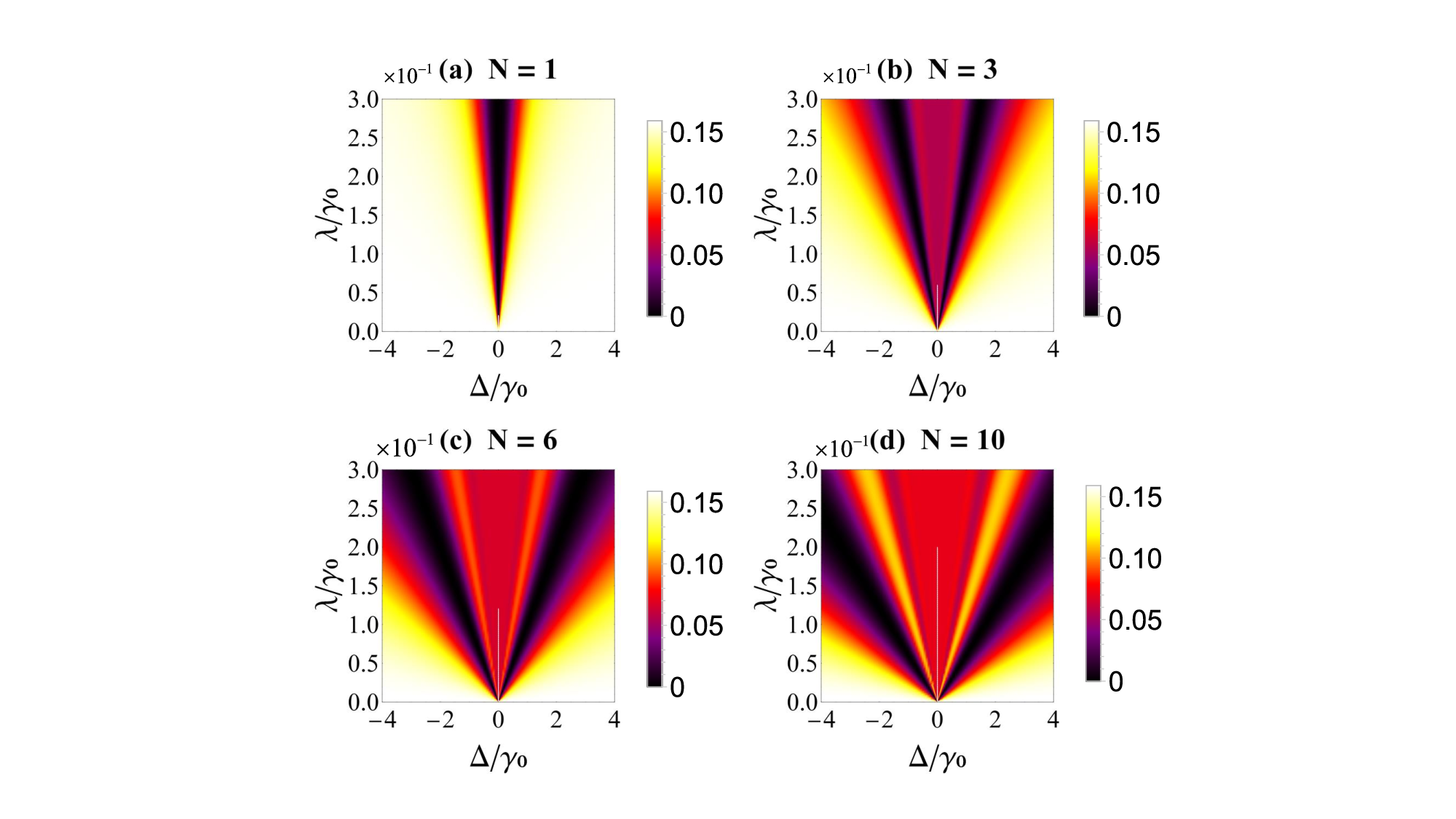}	
	\caption{Maximum synchronization measure $S_m(t)$ as a function of detuning $\Delta$ and spectral width $\lambda$ (in units of $\gamma_0$) at time $\gamma_0t=1000$ for (a) $N=1$, (b) $N=3$, (c) $N=6$, and (d) $N=10$.} 
	\label{Fig5}
\end{figure}

Figure \ref{Fig4} displays the phase synchronization measure ($S_m(t)$) as a function of coupling strength ($\gamma$) and detuning parameter ($\Delta$) for different numbers of auxiliary qubits ($N$) at $\gamma_0t = 1000$ in the non-Markovian regime ($\lambda = 0.01\gamma_0$). Here $\gamma$ denotes a variable coupling parameter expressed in units of $\gamma_0$. To vary the coupling strength, the spectral density $J(\omega)$ is rewritten in terms of this parameter. For a single qubit ($N = 1$; see Fig. \ref{Fig4}(a)), synchronization (indicated by bright regions) occurs outside the expected tongue region, which is centered at $\Delta = 0$. Inside this tongue, the dynamics remain desynchronized. This unexpected result arises from the structured reservoir: at $\Delta = 0$, the qubit is in resonance with the Lorentzian peak, leading to maximum dissipative broadening that overwhelms any potential phase locking. When there is finite detuning, the qubit moves off-resonance, reducing the effective decay rate and allowing the memory kernel to sustain coherent oscillations. This enables synchronization at the detuned frequency. As the number of qubits ($N$) increases (as shown in Fig. \ref{Fig4}(b)–(d)), the collective coupling represented by $N\gamma_0$ alters the effective coupling between the system and its environment. This pushes the system further into the non-Markovian regime and broadens the tongue width. Synchronization regions begin to appear within the tongue as well, because the enhanced memory strength can now support phase locking even at resonance. Consequently, the central line at $\Delta = 0$ signifies a transition from dynamics dominated by dissipation to those dominated by memory, with the additional qubits amplifying the memory effect.

Figure \ref{Fig5} illustrates $S_m(t)$ as a function of $\Delta$ and $\lambda$, using the same parameters as Figure \ref{Fig4}. It is seen that for $N = 1$ (Fig. \ref{Fig5} (a)), a triangular region for $\lambda$ appears at finite $\Delta$, where synchronization occurs only outside of this region. As the number $N$ increases (Fig. \ref{Fig5} (b)–(d)), the stable region expands toward smaller $\lambda$ and broader $\Delta$. Such a behavior is owing to the collective enhancement of memory (through $N\gamma_0$), which compensates for weaker environmental correlations. This demonstrates that auxiliary qubits act as memory amplifiers, extending the parameter window in which detuning-induced locking is effective.
\\

\subsection{{Bloch Sphere Trajectories and Geometric Interpretation}}

The Bloch sphere provides a crucial geometric framework for visualizing the dynamics of a two-level quantum system. In this representation, the reduced density matrix $\rho_i(t)$ of any qubit is expressed as
\begin{equation}
	\rho_i(t) = \frac{1}{2} [I + \mathbf{n}(t) \cdot \boldsymbol{\sigma}],
	\label{Eq20}
\end{equation} 

with $\mathbf{n}(t) = (n_x(t), n_y(t), n_z(t))$ is a real vector constrained to the surface of a unit sphere. By comparing Eq. (\ref{Eq20}) with the reduced density matrix of the system (Eq. (\ref{Eq15})), the Bloch vector components are derived as follows:

\begin{equation}
	\begin{split}
		n_x(t) &= \rho_{10}(0)h(t) + \rho_{01}(0)h^*(t), \\
		n_y(t) &= i[\rho_{10}(0)h(t) - \rho_{01}(0)h^*(t)], \\
		n_z(t) &= 2\rho_{11}(0)|h(t)|^2 - 1.
	\end{split}
\end{equation}

\begin{figure}[h!]
	\centering
	\includegraphics[width=1\textwidth]{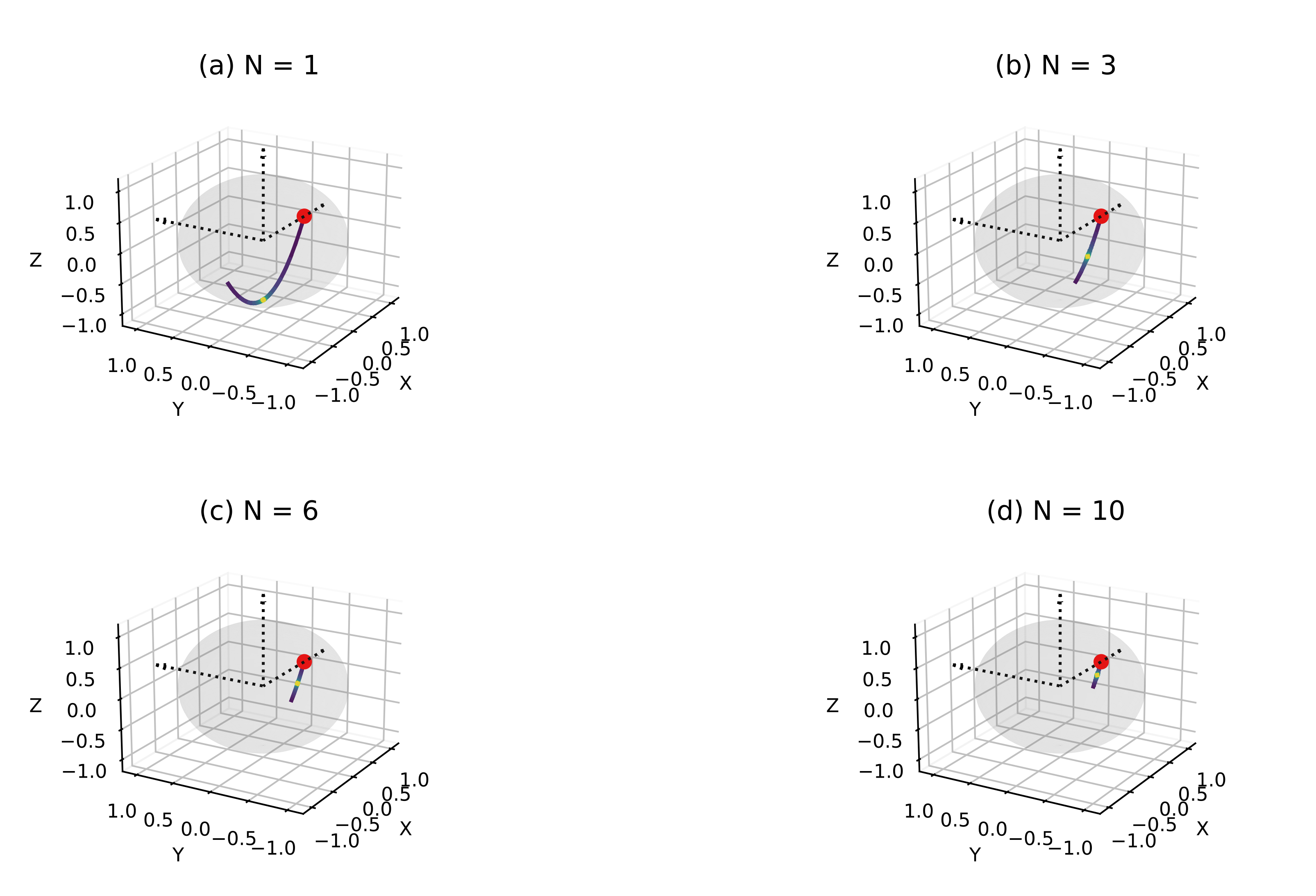}	
	\caption{Bloch-sphere trajectories of the qubit for $\Delta = 0$ up to time $\gamma_0t=1200$ for different numbers of auxiliary qubits: (a) $N=1$, (b) $N=3$, (c) $N=6$, and (d) $N=10$. values of the other parameters are taken as $\lambda=0.01\gamma_0$ and $\gamma = \gamma_0$.} 
	\label{Fig6}
\end{figure}

For pure states, $n(t)$ lies on the surface of the Bloch sphere, whereas for mixed states, it resides within. This representation is particularly valuable in the context of open quantum systems, as it elucidates the complex interplay between coherent and dissipative processes through intuitive trajectories. Unit evolution is represented by rotations of $n(t)$ around a specific axis, while dissipative channels lead $n(t)$ toward the interior or poles of the sphere. In terms of synchronization, the Bloch sphere helps determine whether the system's dynamics converge to a static fixed point (decoherence) or exhibit persistent, periodic orbits (limit cycles). The latter represents the geometric signature of sustained phase locking.
In the following, we will track $n(t)$ to visualize how detuning $\Delta$ induces coherent rotation about the $z$-axis, how environmental memory produces recurrent oscillations through information backflow, and how auxiliary qubits $N$ influence the strength of dissipation and the depth of memory. This geometric perspective deepens our understanding of the physical mechanisms that enable the transition from phase diffusion to robust synchronization, thereby complementing the insights provided by the Husimi $Q$-function.

Figure \ref{Fig6} shows the trajectories of qubits in the Bloch sphere up to time $\gamma_0t=1200$ for various numbers of auxiliary qubits with $\Delta = 0$. In the case of a single qubit ($N = 1$), the trajectory is a monotonic spiral to the ground state (south pole), driven by the exponential factor $e^{-\lambda t/2}$ in $h(t)$. For $N = 3$ and $N = 6$, environmental memory causes the trajectory to execute damped loops as coherence is temporarily restored. The loops shrink with increasing $N$ because the enhanced collective coupling accelerates decoherence of the decaying component, leaving only the protected $\frac{(N-1)}{N}$ fraction in the steady state. For $N = 10$, the trajectory nearly collapses to a fixed point near the initial state ($n_x = 1$), indicating that the strong non-Markovian coupling induces a measurement-like continuous projection onto the decoherence-free subspace—an adequate quantum Zeno effect. No limit cycle forms because, without detuning, there is no energy input to sustain perpetual oscillations. Figure \ref{Fig6} clearly illustrates the dynamics of pure dephasing.

\begin{figure}[h!]
	\centering
	\includegraphics[width=1\textwidth]{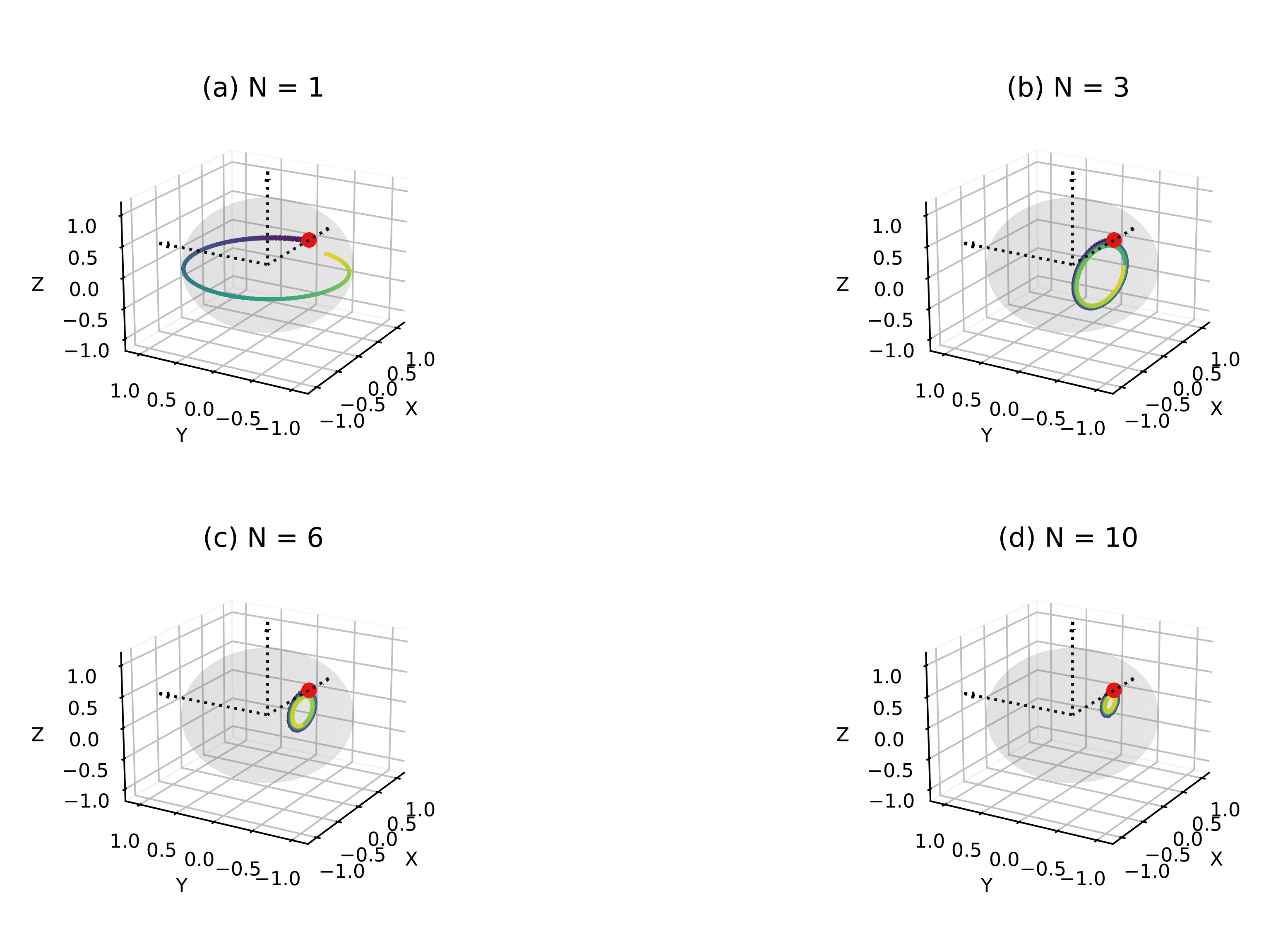}	
	\caption{Bloch-sphere trajectories for $\Delta = \gamma_0$ sing the same parameters as Figure \ref{Fig6}.} 
	\label{Fig7}
\end{figure}

Figure \ref{Fig7} displays the trajectories of the qubit on the Bloch sphere for $\Delta = \gamma_0$, with all other parameters maintained as in Figure \ref{Fig6}. It is clear that the dynamics exhibit a coherent rotation in the presence of detuning. For $N = 1$ (Fig. \ref{Fig7}(a)), this rotation is too weak to overcome dissipation, causing the trajectory to spiral toward the south pole, although with additional twisting. In the case of $N = 3$ (Fig. \ref{Fig7}(b)), the dynamics become more complex: memory backflow periodically reintroduces coherence that is then rotated by the detuning, resulting in a transient limit-cycle-like orbit near the equator. This orbit is quasi-stable because each revival is phase-shifted by the detuning, allowing the system to re-synchronize with its past state after a characteristic memory time $\tau_B$. With $N = 6$ (Fig. \ref{Fig7}(c)), the orbit tightens and lasts longer as memory effects strengthen, demonstrating robust phase locking to the rotating frame induced by detuning. For $N = 10$ (Fig. \ref{Fig7}(d)), the trajectory becomes locked near the initial state. This "freezing" is not due to a suppression of decoherence but rather because the ultrastrong collective coupling causes the environment to function as a continuous, non-destructive measurement apparatus. As a result, the quantum Zeno effect effectively pins the system. This "freezing" is distinct from synchronization; it represents state protection rather than sustained coherent evolution. This highlights that optimal synchronization requires a balance of $N$ to enhance memory without inducing Zeno suppression.

\section{CONCLUSION}
\label{sec.iv}

In summary, we have investigated the emergence of phase synchronization in a single-qubit interacting with a zero-temperature reservoir through the addition of non-interacting auxiliary qubits. While earlier studies emphasized the need for many auxiliary qubits to preserve the initial phase under resonance ($\Delta = 0$), our results reveal that non-zero detuning provides a more powerful and flexible control mechanism, particularly when the system interacts with a non-Markovian reservoir.

It has been demonstrated that detuning and auxiliary qubits act in concert to stabilize quantum phase dynamics. Whereas detuning alone has little influence in the Markovian regime, in non-Markovian environments it interacts constructively with memory-induced revivals, converting them into persistent phase-locked behavior. Auxiliary qubits further reinforce this mechanism by enhancing collective coupling and deepening memory effects, enabling long-lived synchronization with a minimal number of auxiliary qubits. By studying the behavior of the Husimi Q-function, synchronization measure, and Arnold tongues collectively, it was revealed that optimal synchronization arises from a balance between detuning-driven coherent rotation and the memory amplification provided by the auxiliary qubits, establishing a versatile and efficient strategy for engineered quantum phase control. Moreover, the Bloch sphere trajectories highlight that too many qubits push the system toward Zeno-type freezing, underscoring that optimal synchronization arises from a balanced combination of memory amplification and detuning.

Overall, our findings show that the combined use of detuning and auxiliary qubits forms a hybrid and resource-efficient strategy for engineered quantum phase control. This cooperative mechanism provides robust phase stability and is well-suited for applications in quantum communications, precision metrology, phase-synchronized quantum sensing, and synchronization-assisted quantum energy technologies.

\bibliography{mybibb.bib}
\end{document}